\documentclass[preprint]{elsarticle}
\usepackage{amsmath}
\usepackage{amsthm}
\usepackage{textcomp}
\usepackage{lineno}
\usepackage{subfigure}
\usepackage[colorlinks,linkcolor=red]{hyperref}
\journal{Journal of NIM A}
\begin{document}
\begin{frontmatter}
\title{A new layout about the graphite layers in RPC}

\author[keylab,USTC]{Quanyin Li}
\ead{quanyin.li@cern.ch}


\author[keylab,USTC]{Xiangyu Xie}
\author[keylab,USTC]{Zhengwei Xue}
\author[keylab,USTC]{Jiajin Ge}

\author[keylab,USTC]{Yongjie Sun\corref{cor1}}
\ead{sunday@ustc.edu.cn}

\address[keylab]{State Key Laboratory of Particle Detection and Electronics, University of Science and Technology of China, Hefei 230026, China}
\address[USTC]{Department of Modern Physics, University of Science and Technology of China, Hefei 230026, China}
\cortext[cor1]{Corresponding author. Email sunday@ustc.edu.cn; Address:Department of Modern Physics, University of Science and Technology of China, Hefei 230026, China}
\begin{abstract}
The Resistive Plate Chamber (RPC) is widely used in experiments of high energy physics 
as trigger detector as its good time resolution and high efficiency. In the 
traditional layout of RPC, the graphite layers are indispensable parts. 
The working voltage is applied on these layers and the charge of avalanche dissipates 
through them. In this paper, a new design 
which removes the graphate layers is proposed to improve the structure 
of this detector. With this new design, the negative effect from the ununiformity of 
graphite is eliminated and the structure of detector is simplified.

\end{abstract}
\begin{keyword}
    Gaseous detector \sep structure improvement  \sep Graphite layer 
    
\end{keyword}
\end{frontmatter}
\clearpage
\section{Introduction}
\label{sec:intro}
\begin{figure}[!htb]
    \centering
    \includegraphics[width = 1.0\textwidth]{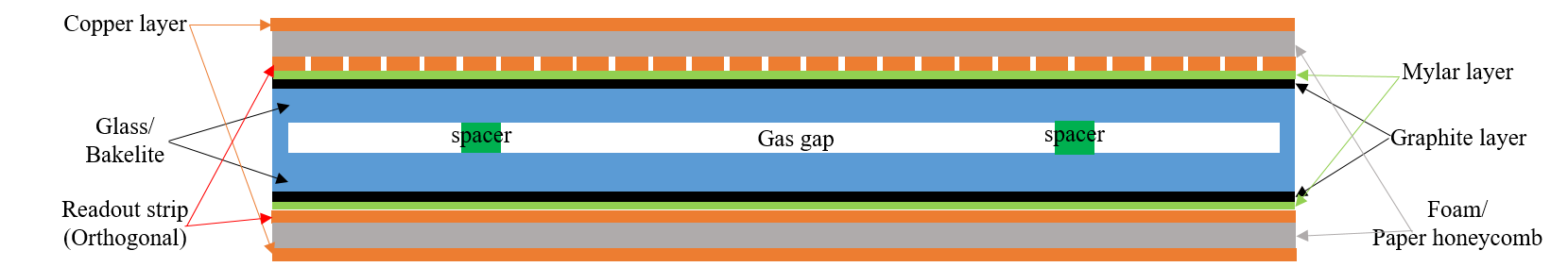}
    \caption{Traditional structure of RPC detector.}
    \label{fig:structure}
\end{figure}

RPC is a kind of gaseous detector wihich was invented during 1980s \cite{bib:RPC1} 
\cite{bib:RPC2}. Until now, it has been one of the most widely used detectors in High 
Energy Physics (HEP) experiments. The traditional structure of RPC is shown is Fig. 
\ref{fig:structure}, including 1 gas gap and 2 readout panels. The gas gap consists 
of spacers and resistive plates with graphite and mylar layers coated on the surface. 
Commonly the material of resistive plates is glass ($\rho$ \textasciitilde $10^{12}$ 
$\Omega$$\cdot$cm) or bakelite ($\rho$ \textasciitilde $10^{10}$ $\Omega$$\cdot$cm).
Two readout panels with orthogonal strips are assembled on both sides of the gas gap. 

As indispensable components of RPC detectors, the graphite layers play an important 
role. The working voltage is applied on these layers and the charge of avalanche 
also dissipates through them. The surface resistivity of the these layers is an 
important parameter which need to be tuned precisely. If the surface resistivity is 
lower than several hundreds $k\Omega/\qedsymbol$, the induced signals will spread to 
a large area through the graphite layers and result in a large cluster size 
\cite{bib:cluster}. On the other hand, if the surface resistivity is too high, the 
large electric potential 
difference will be formed along the surface of detector. This will reduce the effective 
working voltage in the region far away from the injection point of working voltage 
especially for RPCs with very large size and in the high irradiation conditions. 
In this case, the surface resistivity must be 
controlled in a range from several hundreds $k\Omega/\qedsymbol$ to several $M\Omega/\qedsymbol$. 

The strict requirement of the surface resistivity bings some difficulties in coating 
the graphite layers on the resistive plates. And during the coating, it is also difficult 
to keep the uniformity on the whole surface. This is one of the dominated uncertainties 
in terms of manufacturing technology. In order to improve these disadvantages, the new 
layout which removes the graphite layers is proposed and tested in the paper. What's 
more, this new layout can also simplify the structure of RPC detector. 

\section{Design of new RPC structure}
\label{sec:structure}

\begin{figure}[!htb]
    \centering
    \includegraphics[width = 1.0\textwidth]{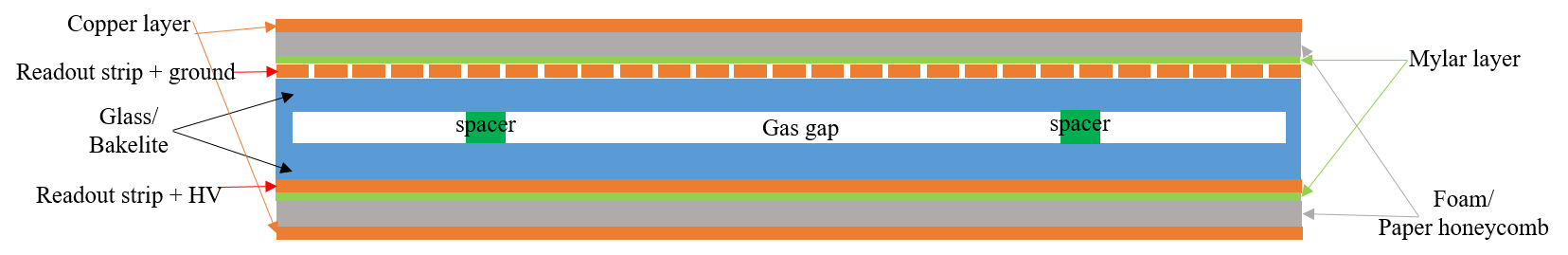}
    \caption{Ideal structure of improved RPC detector.}
    \label{fig:newstructure}
\end{figure}

\begin{figure}[!htb]
    \centering
    \includegraphics[width = 1.0\textwidth]{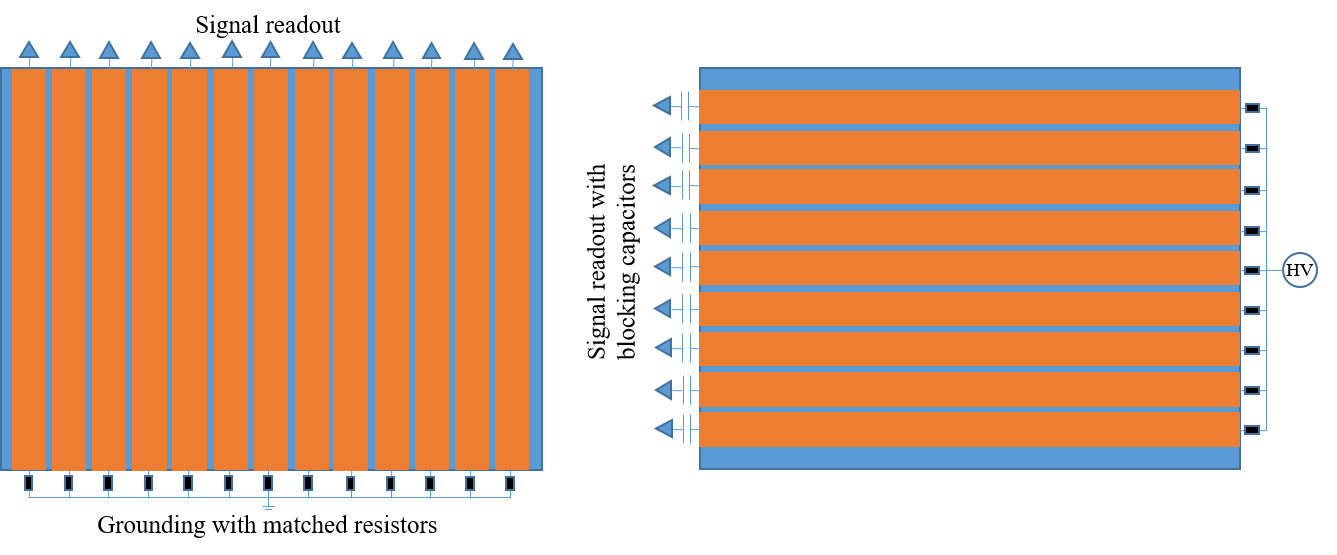}
    \caption{Detailed design of ground strips (left) and HV strips (right) in the improved structure of RPC detector.}
    \label{fig:detailed}
\end{figure}

In the traditional structure of RPC, there are two layers of graphite coated on both 
sides of the gas gap and connected with ground and high voltage (HV) respectively. 
The new design shown in Fig. \ref{fig:newstructure} will remove these two graphite layers. 
In this improved structure, the strips are attached directly to the glass or 
bakelite plates and used for both signal pick-up and voltage applying. 
To achieve 
these goals, as shown in Fig. \ref{fig:detailed}, one end of the strip is 
connected with ground or HV via the matching resistors while the other 
end is connected with front-end electronics for signals pick-up. 

With this new design, in principle, the quality of induced signals is not affected 
or even slightly improved since the strips are closer to the avalanche.
The potential along the surface of each side of the gap is uniform no matter how large 
the detector size is since the strips are made of copper and the matching resistors are 
negligible (\textasciitilde 20 to 30 $\Omega$) compared with the glass or bakelite plates.
Fot the signal readout, the detailed structures for the ground and HV strips are different.
In the HV strips, the blocking capacitors are used to isolate the front-end 
electronics and the power supply to eliminate the latent risk of 
very high potential (\textasciitilde 6000 V for 
RPC with 1 mm gas gap). At the same time, the ground strips are much safer, consequently 
the blocking capacitors could be ignored in the case of high quality of electrical 
grounding. At last, the problem of large cluster size caused by the low surface 
resistivity of the graphite layer can also be eliminated since the signals will not 
spread through this layer.

\section{RPC prototype test}
\label{sec:prototype}
To verify the new structure, two RPC prototypes made of glass plates are produced and 
tested in a cosmic ray telescope. The thickness of gas gaps is 1 mm and the pitch of 
readout unit is 27 mm (25 mm strip + 2 mm gap). The induced signals are read out from 
the ground strips via the homemade amplifiers with gain \textasciitilde 15. Since the 
application of blocking capacitor is a mature technology which has been applied 
successfully in many detectors, such as the ATLAS Monitored Drift Tubes (MDT) 
\cite{bib:MDT}, signals from the HV strips are not 
tested emphatically in this paper.

\begin{figure}[!htb]
    \centering
    \includegraphics[width = 1.0\textwidth]{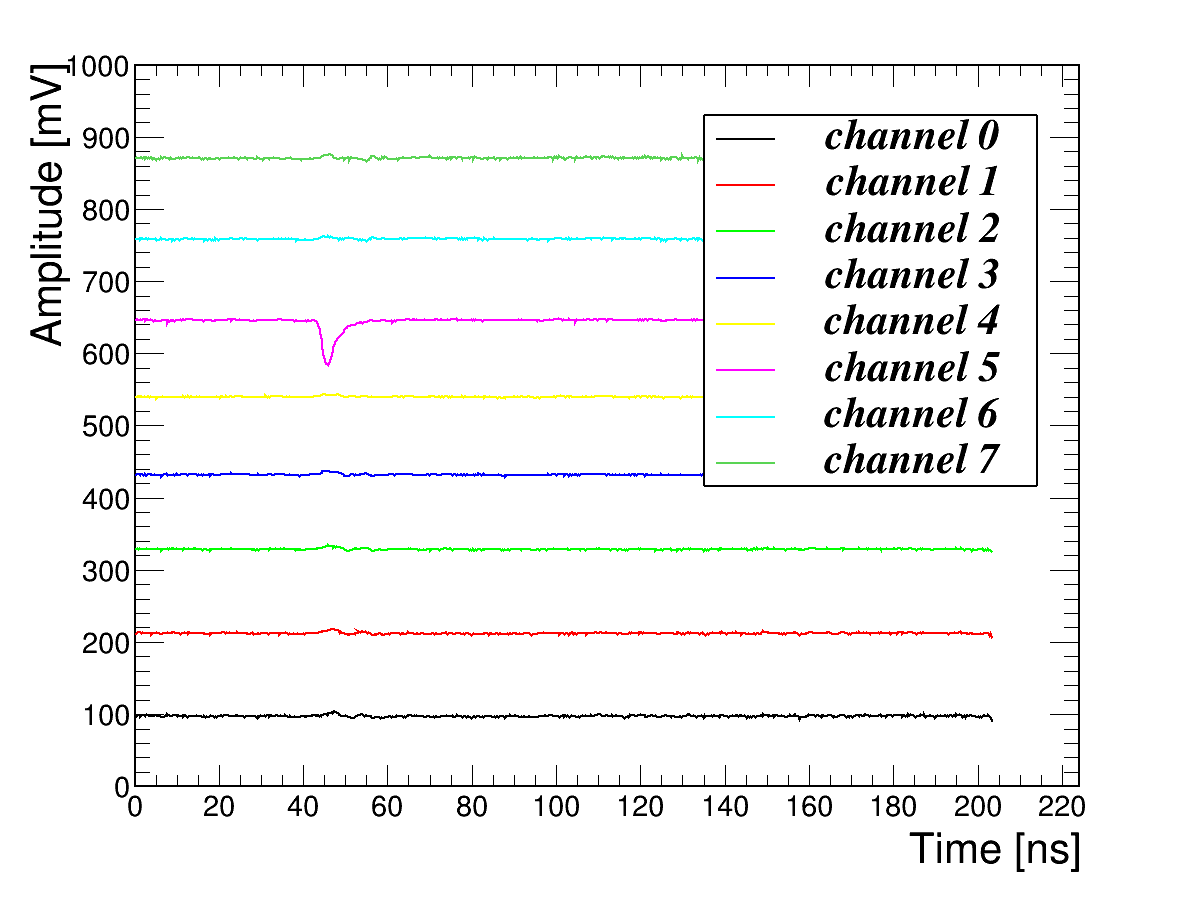}
    \caption{A speciafic example of signal waveform from the RPC with new structure.}
    \label{fig:waveform}
\end{figure}

The classic measurement system built by two RPCs and triggered by the coincidence of 
two scintillators is set up to measure the performance of the 2 RPC prototypes 
\cite{bib:ATLASRPC}. The waveform of signals are recorded completely by the CAEN 
digitizer V1742 with sampling frequency of 5 GS/s. When the working voltage is higher 
than 5600 V, signals with very good quality can be collected by the system. An 
example of signal waveform is shown in Fig. \ref{fig:waveform}. The negative pulse in 
channel 5 is the induced signal. 

\begin{figure}[!htb]
    \centering
    \includegraphics[width = 1.0\textwidth]{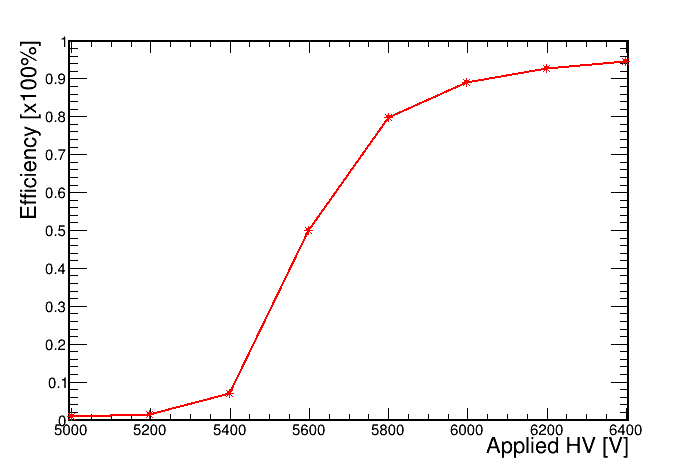}
    \caption{Efficiency plateau of RPC with new structure.}
    \label{fig:eff}
\end{figure}

The efficiency measurement is performed by keeping the HV of one RPC at 6400 V as 
reference and varying that of the other one from 5000 V to 6400 V. A sample of good 
cosmic muons can be selected by requiring the coincidence of the reference RPC and 2 
scintillators. The efficiency is calculated as the ratio of signal number in the tested 
RPC to that in this sample. The efficiency plateau is drawn in Fig. \ref{fig:eff}. 
The maximum at 6400 V is 94.3\% while the minimum at 5000 V is 0.9\%.

\begin{figure}[!htb]
    \centering
    \includegraphics[width = 1.0\textwidth]{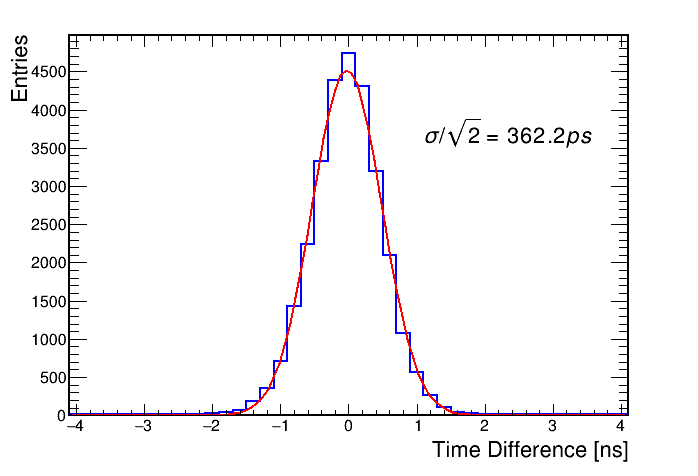}
    \caption{Difference of time-of-arrival between the 2 RPCs.}
    \label{fig:diff}
\end{figure}

The time resolution is measured when both of the 2 RPCs are working at 6400 V. The 
difference of time-of-arrival between the 2 RPCs is shown in Fig. \ref{fig:diff}.
Fit this distribution with a Gaussian function and the sigma divided by sqrt(2) is the 
measured time resolution since the two RPCs are the same. The time resolution is 
measured to be 362.2 ps. 

\begin{figure}[!htb]
    \centering
    \includegraphics[width = 1.0\textwidth]{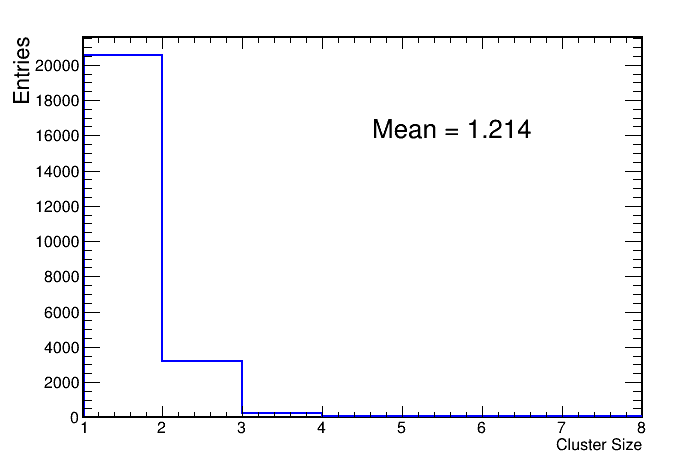}
    \caption{Cluster size of RPC with new structure.}
    \label{fig:cs}
\end{figure}

As the removal of graphite layer, the cluster size is expected to be smaller. The
average value of this variable is measured to be 1.21 and the distribution is shown 
in Fig. \ref{fig:cs}.

Compared with the RPCs with traditional structure \cite{bib:TDR}, our RPC prototypes 
have the similar performance and simplified structure. 

\section{Summary}
\label{sec:summary}
To improve the disadvantage of graphite layers in RPC detector, the RPC prototypes which 
remove these layers and use the strips as both signal pick-up and voltage applying is 
proposed and tested. High efficiency and good time resolution have been measured. 
In the future, this improvement of graphite 
layers could also be attempted in some other gaseous detectors such as MRPC and TGC. 
This will be a progress in eliminating the negative effect from the ununiformity of
graphite layers and simplifyng the structure of some gaseous detectors.

\clearpage
\section*{Acknowledgement}
This study was supported by National Key Programme for S\&T Research and Development 
(Grant NO.: 2016YFA0400100) and the National Natural Science Foundation of China 
(No. 11961141014). This work is also supported by the State Key Laboratory of 
Particle Detection and Electronics, SKLPDE-ZZ-202011 and in part by the CAS 
Center for Excellence in Particle Physics (CCEPP).
\newline

\bibliographystyle{elsarticle-num-names}

\end{document}